\documentclass[%
reprint,
superscriptaddress,
amsmath,amssymb,
pra,
]{revtex4-2}

\usepackage{graphicx}  % needed for figures
\usepackage{dcolumn}   % needed for some tables
\usepackage{bm}        % for math
\usepackage{amssymb}   % for math
\usepackage{amsmath}   % for math
\usepackage{lineno}		% for line numbers
\usepackage[usenames, dvipsnames]{color}
\usepackage{booktabs} 
\usepackage{array}
\usepackage{url}

\graphicspath{{Figures/}} 

\makeatletter
\DeclareGraphicsExtensions{.png, .pdf}
\DeclareRobustCommand{\textsubscript}[1]{{%
  \m@th\ensuremath{%
    _{\mbox{\tiny{#1}}}%
  }%
}}
\makeatother

\begin{document}

\title{Recognition of auditory objects from continuous and ultra-sparse spike streams in neuromorphic devices}

\newcommand{\psu}{Department of Physics and Huck Institutes of the Life Sciences, Pennsylvania State University, University Park, Pennsylvania 16802, USA}

\author{Leonardo E.~S.~Tavares}
\affiliation{\psu}
\author{Kevin D.~Sargent}
\affiliation{\psu}
\author{Jon T.~Sakata}
\affiliation{Department of Biology, McGill University, Montreal, Quebec, Canada}
\author{Dezhe Z.~Jin}
\email[Corresponding author: ]{dzj2@psu.edu}
\affiliation{\psu}

\begin{abstract}
Energy consumption is a major concern for modern artificial intelligence (AI) applications. The brain is thought to achieve energy-efficient computation through sparse coding schemes. By emulating the spiking dynamics of the brain, neuromorphic chips hold the potential to realize energy-efficient AI. In this work, we use birdsong to demonstrate that auditory object recognition can be achieved using ultra-sparse coding. Feature-detecting neurons tuned to short temporal features ($\sim$37 ms), sampled from exemplar syllables, transform continuous auditory signals into streams of ultra-sparse spikes that mark the onsets and offsets of features within the signal. Spike patterns elicited by the exemplars are stored as templates. During recognition, template matching is performed using axonal delays between feature-detecting and template-detecting neurons. Recognition is further enhanced by an additional group of specialized binary neurons trained to discriminate between syllable categories with subtle differences. When tested on Bengalese finch songs, our approach achieved error rates below 5\% using 10 to 40 exemplars per syllable type. We then implemented this recognition algorithm using spiking neural networks on Intel’s neuromorphic chip, Loihi, and found that response latencies remain constant (approximately 17 ms per syllable), regardless of the number of templates used. Our results demonstrate the potential of neuromorphic chips for low-latency, low-energy recognition of complex sounds, including speech.
\end{abstract}

\maketitle

%\linenumbers

\section{Introduction}
\label{sec:intro}

Automatic speech recognition (ASR) is a key technology for human–machine interfaces and speech transcription. Recent advances in artificial intelligence (AI) have significantly improved ASR performance \cite{baevski2020wav2vec,radford2023robust,gulati2020conformer,peng2023reproducing,puvvada2024less,ghosh2026audio}. Large models such as Whisper from OpenAI \cite{radford2023robust}, trained on 680,000 hours of labeled speech collected from the Internet, demonstrate robustness against noise corruption in speech signals \cite{gong2023whisper}.

Nonetheless, the deep learning paradigm on which these systems are built requires a substantial amount of energy for training \cite{schwartz2020green,parcollet2021energy}. For instance, Canary, an ASR model from NVIDIA, was trained on 86,000 hours ($\sim$9.8 years) of audio data using 128 NVIDIA A100 80GB GPUs over a period of 48 hours \cite{puvvada2024less}. We estimate that the energy consumption for this process is on the order of 2 MWh, which corresponds to approximately 20\% of the annual energy consumption of a typical U.S. household. Because hyperparameter tuning requires multiple rounds of training, the energy costs associated with optimizing these models are considerable \cite{schwartz2020green}. Beyond training, the energy cost of inference also increases with model size, which correlates with accuracy \cite{samsi2023words}.

The human brain is remarkably energy-efficient, consuming on the order of 20 W \cite{kovavc201020}. Neuronal activity, including signal transmission, accounts for more than half of the brain’s energy consumption \cite{laughlin2003communication}. This suggests that the brain’s energy efficiency may be due to the use of sparse spike coding \cite{levy1996energy,simoncelli2001natural}. Indeed, neural population responses to natural stimuli are highly sparse in both the visual cortex \cite{tang2018large} and high-level auditory areas \cite{hromadka2008sparse,schneider2013sparse}.

Neuromorphic computing emulates the spiking dynamics of the brain on hardware to achieve energy-efficient, low-latency computation \cite{mead1990neuromorphic,davies2018loihi,furber2020spinnaker,kudithipudi2025neuromorphic}. With millions of spiking neurons implemented on processing chips that can be stacked to scale up network size, neuromorphic platforms such as Loihi 2 have the potential to provide energy-efficient, real-time solutions for a range of image and auditory processing tasks \cite{shrestha2024efficient}. The Loihi platform has supported the development of useful algorithms, such as LASSO regression implemented using spiking dynamics \cite{davies2021advancing}. Loihi chips are also employed in practical applications, including image and speech recognition \cite{stewart2023speech2spikes,shrestha2024efficient}. A common approach to signal decoding is to use spiking versions of deep neural networks trained via backpropagation. However, this approach typically yields less accurate results compared to conventional deep neural networks \cite{davies2021advancing}. Key to the limited practical use of neuromorphic computing is the lack of algorithms that take advantage of the temporal dynamics of spiking neural networks for computation.

In this paper, we show that ultra-sparse coding of auditory signals can be used for the recognition of auditory objects from continuous signals. In particular, we address the task of labeling syllables in continuous birdsong recordings. Automatic labeling of birdsong syllables is important for studying the neural basis of complex vocalizations \cite{doupe1999birdsong,jin2013neural}, and various machine learning techniques have been developed for this purpose \cite{koumura2016automatic,cohen2022automated,vengrovski2025tweetybert}. Birdsong syllables are stereotyped and can be categorized into a small number of types, but still show many features in common with speech \cite{doupe1999birdsong}. Therefore, this task provides an ideal testbed for developing novel speech recognition techniques without confronting the full complexity of human speech.

We train feature-detecting neurons (feature neurons) that respond to complex spatiotemporal acoustic patterns \cite{schafer2014noise}. A collection of feature neurons transforms the continuous auditory signal into a stream of ultra-sparse spikes. By matching the spatiotemporal spike patterns to templates learned for auditory objects, a sequence of labels is produced. We implement the template matching process on both a CPU and Loihi and show that the response latency on Loihi remains constant as the number of templates increases, while on the CPU it increases superlinearly. Our results demonstrate that neuromorphic chips such as Loihi can use ultra-sparse spike coding for pattern recognition tasks and hold potential for achieving low latency and low energy consumption in applications such as speech recognition.

\section{Results} 
\label{sec:results}

\subsection{Encoding syllables into ultra-sparse spike sequences}

\begin{figure*}[!tp]
	\includegraphics[width=\textwidth]{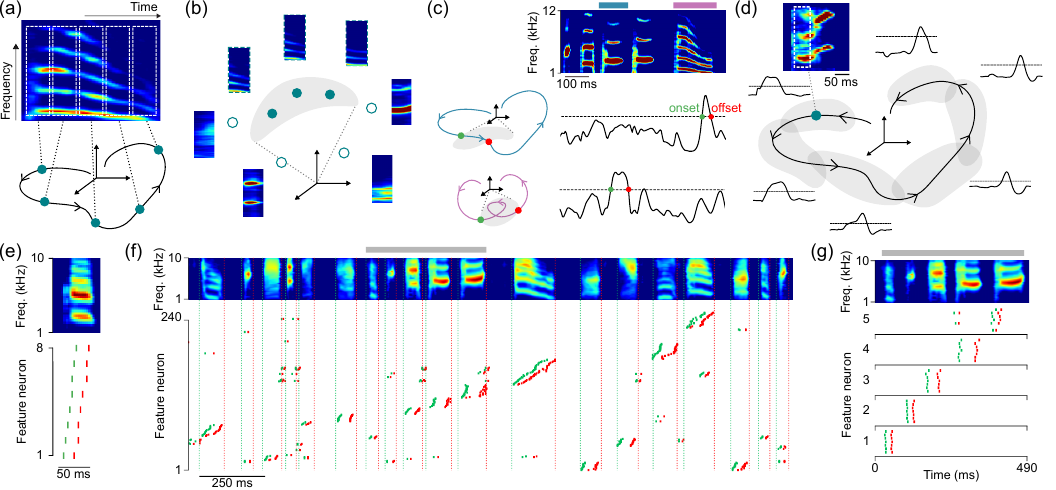}
\caption{{\bf Encoding birdsong syllables into ultra-sparse spike sequences}. (a) Overlapping windows (dashed boxes) slide along the spectrogram of a Bengalese finch syllable (top). Each window is a feature vector, i.e., a point in feature space. The syllable traces out a trajectory (bottom). (b) Feature vectors from the same part of one syllable type (filled circles) lie close together in the feature space and occupy a volume (gray) that can be separated from feature vectors of other syllable types (open circles). A feature neuron learns the boundary of this volume. (c) Responses of two feature neurons (right) to continuous song (top). Left: the trajectories of the two syllables and the decision boundaries (gray) of the two neurons. When the trajectory enters the decision boundary, the response crosses a threshold (dashed lines) and the neuron fires an onset spike (green); when it leaves, the response falls below threshold and the neuron fires an offset spike (red). A sustained feature produces a plateau in the response (bottom). (d) A chain of decision boundaries, one per feature neuron, covers the entire trajectory of a syllable; the traces show the responses of the corresponding feature neurons. The syllable is thus represented as a sequence of discrete events. (e) Onset (green) and offset (red) spikes of feature neurons evoked by an isolated syllable. (f) Onset and offset spikes of the feature-neuron population evoked by a continuous song. Green and red dotted lines mark the syllable onsets and offsets determined from the spectrogram. (g) Onset and offset spikes of 5 feature neurons evoked by 7 renditions of the song segment marked by the gray bar in (f). The spikes are reliable and temporally precise.}
\label{Figure1}
\end{figure*}

Song recordings were manually annotated and converted into spectrograms using fourth-order gammatone filters that simulate the spectral analysis performed by the cochlea \cite{patterson1987efficient} (Appendix~\ref{sec:methods:feature_processing}). Spectrogram segments were then mapped into vectors in a feature space [Fig.~\ref{Figure1}(a)]. Specifically, feature vectors were constructed by concatenating 11 spectrogram frames within a $\sim$37 ms temporal window, each frame representing $3.3$ ms and spanning a frequency range of 1--10 kHz in 32 bands, resulting in a 352-dimensional feature space. As the window advanced in time, syllables were mapped into continuous trajectories in this space [Fig.~\ref{Figure1}(a)].

We trained feature-detecting neurons (feature neurons) to respond to acoustic features shared by syllables of the same type. For each type, we first drew at random a reference syllable. Next, for each feature vector in the reference, we calculated from the other syllables of the same type the vectors that correlated the highest with the reference vector. This produced a cluster of vectors representing the same part of the syllable across renditions (Appendix~\ref{section:methods:feature_neurons}). Using support vector machines \cite{cortes1995support} (SVMs), we trained feature neurons to discriminate the vectors of each cluster against other acoustic events, including other syllable types, silence, and noise [Fig.~\ref{Figure1}(b); Appendix~\ref{section:methods:feature_neurons}].

Mathematically, feature neurons are defined by the SVM decision function
\begin{equation}
V(t) = \sum_{i=1}^n \alpha_i y_i \exp \left(-\gamma \lVert {\bf x}(t) - {\bf x}_i \rVert^2 \right) + V_0.
\label{equation:feature_detector_response}
\end{equation}
Here, ${\bf x}(t)$ is the input feature vector; ${\bf x}_i$ are the $n$ support vectors; $y_i$ equals 1 for support vectors in the target syllable type (positive feature vectors) and $-1$ for those in the background (negative feature vectors); $\alpha_i$ and $V_0$ are the learned parameters and $\gamma$ is calculated from the training data (Appendix~\ref{section:methods:feature_neurons}). The input feature vectors ${\bf x}(t)$ are parameterized by time, and the parameter $\gamma$ controls the smoothness of the decision boundary (by definition, the collection of points satisfying $V(t)=0$ (Appendix~\ref{section:methods:feature_neurons}). We interpret $V(t)$ as the “membrane potential” of the feature neuron, by analogy with biological neurons.

The membrane potential fluctuates below threshold ($V(t)=0$) for most inputs. It reliably crosses the threshold when the neuron's positive feature appears in the input and eventually returns to baseline after the positive feature passes. Two examples are shown in Fig.~\ref{Figure1}(c). We defined and recorded threshold crossings from below as onset spikes (green dots in Fig.~\ref{Figure1}(c); onset for short) and crossings from above as offset spikes (red dots in Fig.~\ref{Figure1}(c); offset for short). A sustained feature, such as a flat harmonic stack, can produce a plateau in the membrane potential [Fig.~\ref{Figure1}(c), bottom panel]. Thus, feature neurons can respond to sustained features of variable duration.

By definition, a syllable consists of a sequence of feature vectors. If a collection of decision boundaries is trained to cover this sequence, these feature vectors will enter and leave the boundaries orderly [Fig.~\ref{Figure1}(d)], effectively creating an ultra-sparse sequence of onset and offset spikes. Extending this idea to all types enables each syllable to elicit a distinctive spike sequence in a subset of feature neurons that also identifies that syllable's type. Thus, syllable types are encoded by ultra-sparse spike patterns produced by the population of feature neurons [Fig.~\ref{Figure1}(e)]. 

Collectively, feature neurons transform continuous audio inputs into ultra-sparse onset and offset spikes [Fig.~\ref{Figure1}(f)]. As feature neurons spike only in response to their positive features, this sparseness reflects the prevalence of that feature in the input sound. These spike patterns are highly stable across renditions of the same syllable [Fig.~\ref{Figure1}(g)], although missing or erroneous spikes can occasionally occur due to variability in syllable acoustics [feature neuron 5 in Fig.~\ref{Figure1}(g)].

\subsection{Few-shot learning of spike templates}

We collected $N = 5, 10, 20, 40$ syllables per type. For each syllable, we recorded and stored as templates the onset and offset times of feature neurons that spiked. Thus, each syllable type had in total $N$ templates. Increasing $N$ allowed the set of templates to capture more acoustic variability of the syllable categories. This can be viewed as a form of few-shot learning or memorization of spike patterns. The ultra-sparseness and robustness of the responses of feature neurons ensure that templates remain distinctive across different syllable types.

\subsection{Decoding syllables from continuous ultra-sparse spike sequences}

\begin{figure*}[!tp]
\includegraphics[width=\textwidth]{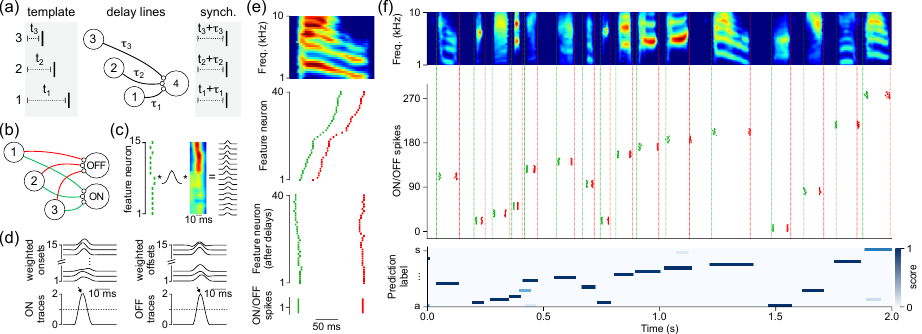}
\caption{{\bf Segmenting and classifying syllables in a continuous recording using delay lines}. (a) Principle of delay-line template matching. Left: a template in which neurons 1--3 spike at times $t_1$, $t_2$, and $t_3$. Middle: the neurons project to neuron 4 with delays $\tau_1$, $\tau_2$, and $\tau_3$. Right: the delays compensate for the differences in spike times, so the spikes arrive at neuron 4 simultaneously ($t_1+\tau_1=t_2+\tau_2=t_3+\tau_3$) and are easily detected. (b) Separate delay lines carry the onset (green) and offset (red) spikes of feature neurons to an ON and an OFF template neuron. (c) The synchronized spikes (left) are convolved with a Gaussian kernel to tolerate timing jitter and filtered with learned weights (color map), so that reliably firing feature neurons contribute more (right). (d) The weighted onsets and offsets (top) are summed into ON and OFF traces (bottom); ON and OFF spikes are emitted at peaks (arrows) above threshold (dotted lines). (e) Example for one syllable: spectrogram (top), onset (green) and offset (red) spikes of feature neurons (2nd panel), the same spikes after the delay lines (3rd panel), and the resulting ON and OFF spikes (bottom). (f) A continuous song (top) and the ON (green) and OFF (red) spikes of all template neurons, ordered by neuron ID, of a system trained on $N=15$ syllables per type (middle). Green and red dotted lines mark the syllable onsets and offsets determined from the spectrogram. Pairing ON and OFF spikes yields the segmentation (bottom): each bar is a predicted syllable, labeled a--s and shaded by its score.}
\label{Figure2} 
\end{figure*}

To recognize (detect and classify) syllable types from continuous spike streams, we employed a template matching scheme based on the expected relative spike times of feature neurons. In this scheme, neuronal connections were assigned delays that compensated for the typical arrival time of spikes [Fig.~\ref{Figure2}(a)], ensuring that the target neuron (the neuron receiving the connections) received all feature neuron spikes synchronously [Fig.~\ref{Figure2}(a)]. Two template-decoding target neurons (template neurons for short) were assigned per template, one for onsets and another for offsets [Fig.~\ref{Figure2}(b)], each receiving the spikes of the feature neurons in that template. Thus, templates were directly encoded in the delay lines between feature neurons and template neurons.

To accommodate variability in spike timing, we smeared spikes with a Gaussian kernel of standard deviation 5 ms and passed the results through a set of trained linear filters (Appendix~\ref{sec:methods:template_matching}). This effectively weighted the contribution of each neuron so that those that fired more reliably had higher weights [Fig.~\ref{Figure2}(c); Appendix~\ref{sec:methods:template_training}]. When the sum of these responses exceeded a threshold, the template neuron fired, indicating recognition of the input spike pattern [Fig.~\ref{Figure2}(d)]. From here on, we refer to template neurons as ON and OFF neurons (depending on whether their inputs are onsets or offsets, respectively) and to the onset and offset spikes they produce as ON and OFF spikes. Thresholds and filters were optimized to maximize the response of the neuron to its assigned syllable type while minimizing its responses to other types [Fig.~\ref{Figure2}(d); Appendix~\ref{sec:methods:template_training}]. 

A pair of ON and OFF spikes indicates a positive match to a template [Fig.~\ref{Figure2}(e)]. Since there are multiple templates per syllable type, multiple detections can be made simultaneously, conveying redundant information [Fig.~\ref{Figure2}(f)]. For any group of overlapping detection, we kept the one with the highest combined ON and OFF responses (Appendix~\ref{sec:methods:template_matching}). This selection provided both the detection and the boundaries of the predicted syllables [Fig.~\ref{Figure2}(f)].

\subsection{Binary discrimination}

\begin{figure}[t!]
\begin{center}
\includegraphics[width=\columnwidth]{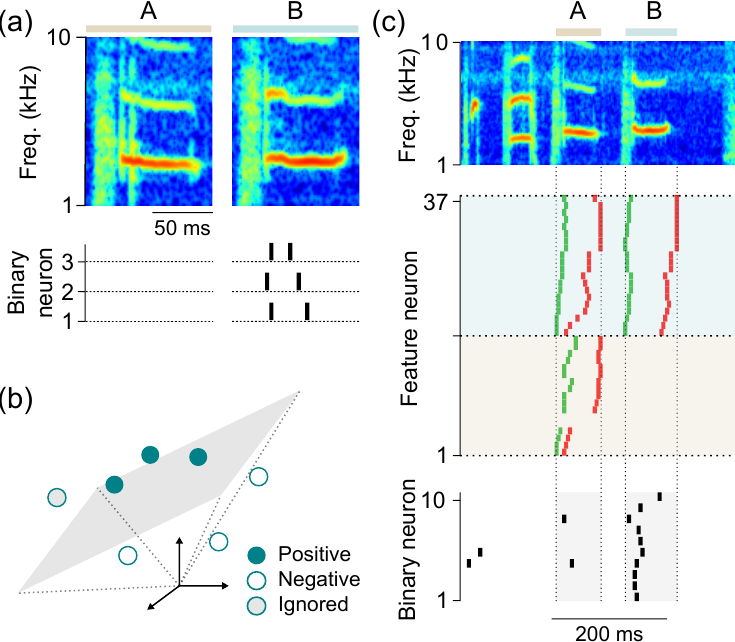}
\end{center}
\caption{{\bf Binary neurons distinguish highly similar syllables}. (a) Two syllables, A and B, with similar acoustic features. Binary neurons trained to separate B from A are silent during A (bottom left) but respond to features of B that are absent from A (bottom right). (b) Diagram of binary-neuron classification. A binary neuron separates feature vectors of one syllable type (positive, filled circles) from those of another (negative, open circles) and ignores vectors from all other types (gray circles), so a simple linear boundary suffices. (c) Continuous recognition with binary neurons. Feature neurons tuned to A (beige) and to B (blue) both respond during syllable A (middle panel; green, onset spikes; red, offset spikes), so feature neurons alone cannot distinguish the two syllables. Binary-neuron spikes are read out only within segments delimited by ON and OFF spikes (gray boxes; dotted lines). Within these segments, the binary-neuron ensemble responds robustly to B and weakly to A (bottom).}
\label{Figure3} 
\end{figure}

Some syllable types can exhibit subtle differences. As illustrated in Fig.~\ref{Figure3}(a), syllables A and B appear very similar, but the harmonic stack in A has a slight pitch downshift, whereas B has a constant pitch. Due to their high similarity, feature neurons of both types tend to spike simultaneously in response to either syllable. In addition, a syllable may sometimes be a segment of a longer one. In this case, feature neurons tuned for the shorter syllable will respond within the duration of the longer one, leading to conflicting recognition labels.

We resolved this issue by training binary neurons to discriminate features of type B from features of type A [Figs.~\ref{Figure3}(a)–\ref{Figure3}(b)]. Each binary neuron examined a small spectrotemporal patch sampled at a random position in the spectrogram, lasting between 13 and 33 ms and spanning at least 8 of the 32 frequency bands. Binary neurons were linear classifiers trained to distinguish the patch in syllables of type B from the same patch in syllables of type A. We used AdaBoost \cite{freund1997decision} to select binary neurons one at a time and weight their contributions, producing an ensemble of 10 binary neurons for per ordered pair of syllable types (Appendix~\ref{sec:methods:binary_neurons}). To speed up training and recognition, ensembles were trained only for pairs of syllable types that produced detection collisions during recognition (Appendix~\ref{sec:methods:template_matching}).

Binary neurons are informative only when conditioned on one or both of the syllable types to which they were trained to discriminate. The same binary neurons are ineffective when applied to other categories or when used in isolation. In Fig.~\ref{Figure3}(c), we show the responses of both feature neurons and binary neurons to a sequence of syllables. A large fraction of the feature neurons spike during both syllables A and B, making it difficult to distinguish between the two. In contrast, binary neurons trained to distinguish B from A spike vigorously during syllable B but only weakly during syllable A. Thus, binary neurons provide additional discriminative power for distinguishing syllable type B from type A.

\subsection{Continuous recognition on CPU}

\begin{figure*}[!tp]
\includegraphics[width=\textwidth]{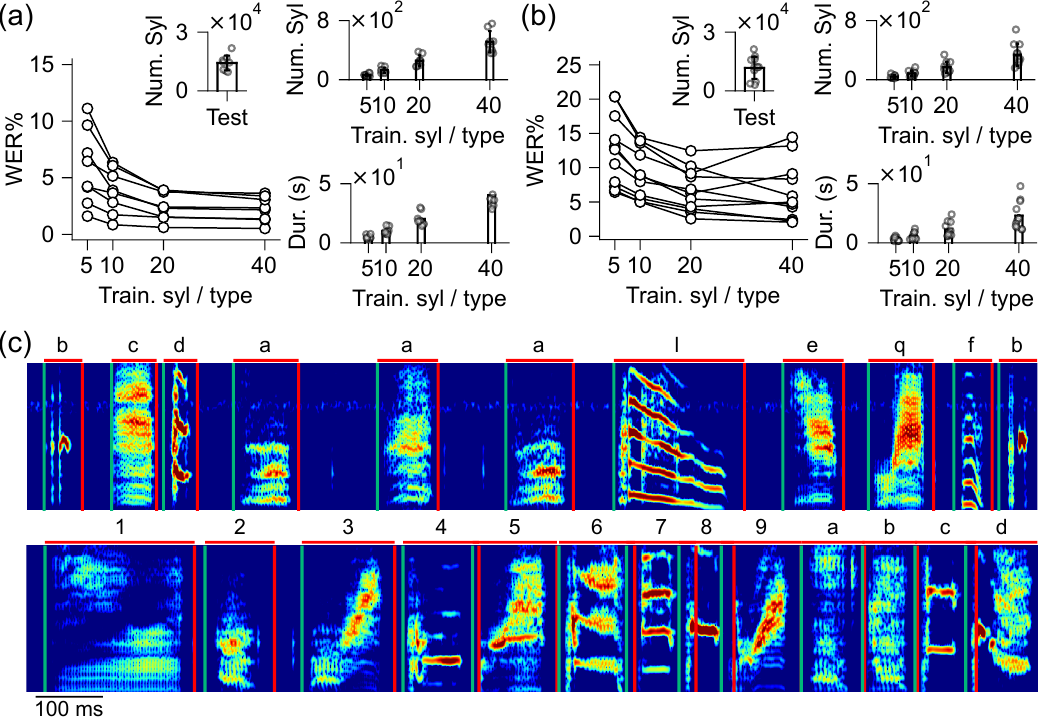}
\caption{{\bf Continuous recognition of birdsong syllables on the CPU}. (a), (b) Recognition results for two benchmarks: 8 Bengalese finches from this paper (a) and 11 Bengalese finches from the uncurated dataset of Ref.~\protect\cite{koumura2016automatic} (b). Left: word error rates (WERs) versus the number of training syllables per type (5, 10, 20, or 40). Each line is one bird; each point is the average over 10 systems trained on independently sampled training sets (see Appendix~\ref{sec:methods:benchmark}). Inset: total number of test syllables. Right: total number of training syllables (top) and their total duration in seconds (bottom). Bars show mean $\pm$ SD across birds (dots). (c) Examples of recognition output for a bird from benchmark (a) (top) and from benchmark (b) (bottom). Green and red lines mark the start and stop of each recognized syllable, and the labels above give the recognized syllable types.}
\label{Figure4} 
\end{figure*}

We evaluated the performance of our recognition algorithm using our recordings of 8 Bengalese finch songs \cite{james2020statistical} (Appendix~\ref{sec:datasets}) and a public dataset of an additional 11 Bengalese finches \cite{koumura2016automatic}. Recordings were made in isolated soundproof chambers \cite{james2020statistical}, which reduced background noise but did not completely eliminate non-vocal sounds such as perching and wing flaps. Our recordings also included calls, which are not part of the songs. We manually annotated all of our recordings to ensure that syllables were assigned the correct categories and boundaries were precise, but did not curate the data of Ref.~\cite{koumura2016automatic}. Recordings were split 60\% for testing and the remaining 40\% to sample training sets. For each bird, we randomly selected $N$ exemplars per syllable type from the training recordings, with $N = 5, 10, 20, 40$. The total number of training syllables and the total duration varied between birds due to differences in their vocal repertoires [Figs.~\ref{Figure4}(a)-\ref{Figure4}(b), right panels].

Performance on test sets was quantified using the word error rate (WER), computed as
\begin{equation}
{\rm WER} = \frac{S + D + I}{K}.
\label{equation:wer}
\end{equation}
Here, $K$ is the total number of annotated syllables, $S$ is the number of substitution errors (syllables incorrectly labeled), $D$ is the number of deletion errors (syllables for which no label was generated), and $I$ is the number of insertion errors (labels generated where no syllables were present). WERs rapidly converged to stable values by 20 exemplars per syllable type for most birds, with the majority averaging below 5\% [Fig.~\ref{Figure4}(a)]. Syllable boundaries were also largely correct [Fig.~\ref{Figure4}(c)].

\subsection{Continuous recognition on the neuromorphic device Loihi}

\begin{figure*}[!tp]
\includegraphics[width=0.95\textwidth]{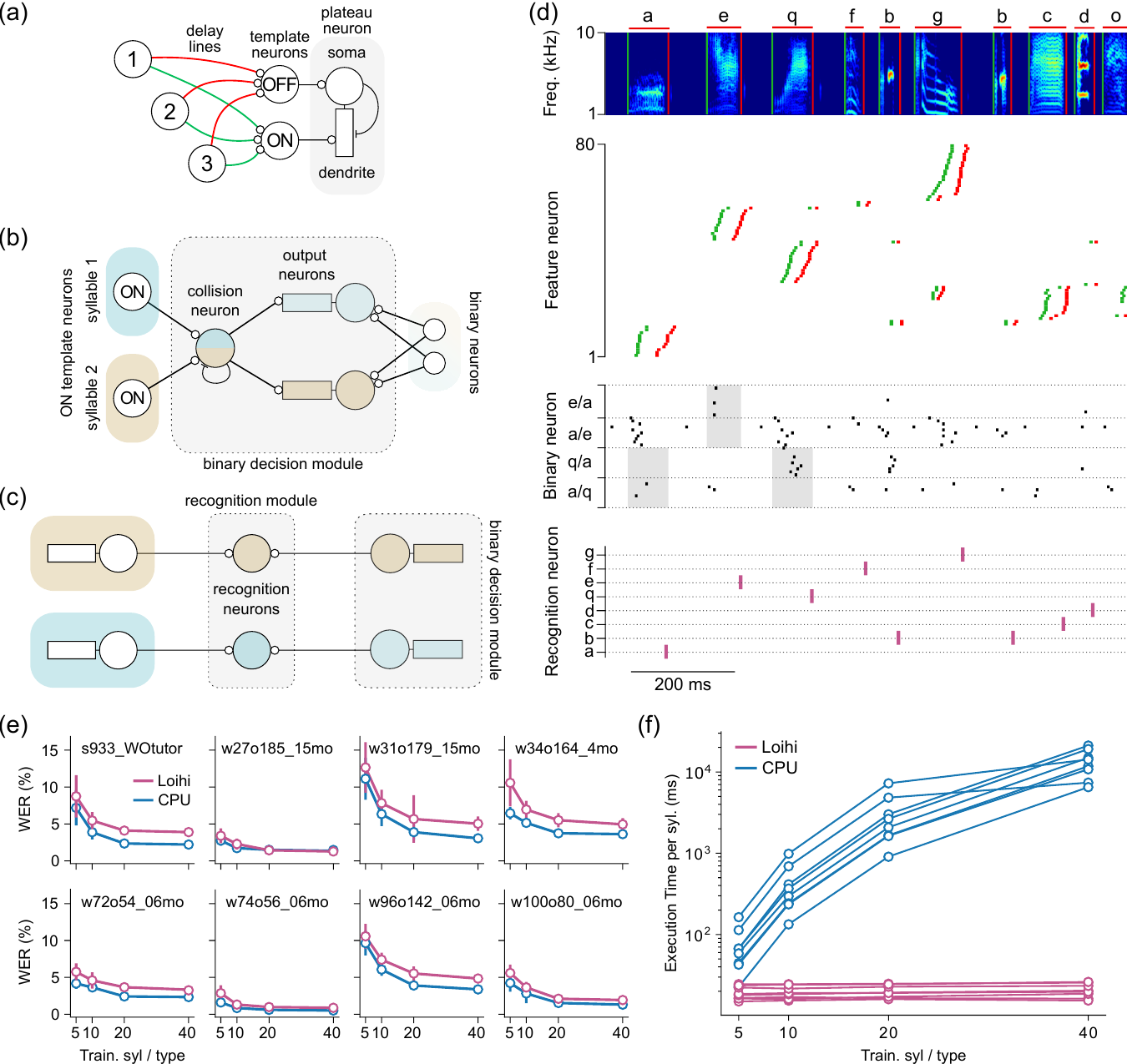}
\caption{
{\bf Continuous recognition of birdsong syllables on neuromorphic hardware}. (a) Diagram of the template-matching module, shown for a single template.
ON and OFF template neurons, which are single-compartment leaky integrate-and-fire (LIF) neurons, receive the onset (green) and offset (red) spikes of feature neurons (numbered circles), respectively, through delay lines that align the feature-neuron spikes so that their coincident arrival can be detected.
The plateau neuron (gray box), a two-compartment LIF neuron shared by all templates of a syllable type, detects ON spikes followed by OFF spikes.
ON spikes arriving at the dendrite raise its voltage above threshold, putting the neuron in the UP state (plateau).
If OFF spikes then arrive at the soma, the soma emits a spike, which inhibits the dendrite and terminates the plateau (see Appendix~\ref{sec:methods:plateau_neurons}).
Without a plateau, input to the soma alone cannot make it spike.
A somatic spike signals detection of the syllable type.
(b) Diagram of the binary decision module. A collision neuron receives spikes, relayed through one coincidence-detection neuron per syllable, from the ON template neurons of two syllables, e.g., syllables 1 and 2, and signals when both syllables are detected at the same time; its inhibitory self-connection prevents repeated spikes.
Collisions are resolved by two-compartment output neurons.
Spikes of the collision neuron arrive at the dendritic compartments of the output neurons and create plateau potentials; if the somatic compartments then receive spikes from the binary neurons, the output neurons spike.
(c) Diagram of the recognition module. Plateau neurons (left) and the output neurons of the binary decision module (right) drive the recognition neurons (LIF neurons), which carry the final decision.
(d) Sample input and network response. The spectrogram (top) is converted into onset and offset spikes (2nd panel) and binary-neuron spikes (3rd panel), which are fed into the Loihi network simultaneously. The network outputs recognition-neuron spikes (4th panel) that predict syllable categories and endpoints. Binary neurons are informative only when one of the two syllables they were trained on is present (gray boxes in the 3rd panel).
(e) WERs on the CPU [blue; reproduced from Fig.~\ref{Figure4}(a)] and on Loihi (magenta), shown as mean $\pm$ SD.
(f) Average execution time per syllable on the CPU (blue) and on Loihi (magenta). Each line is one bird from our dataset.
}
\label{Figure5} 
\end{figure*}

If all templates operate in parallel, the response latency should remain constant despite increases in the number of templates. Such extreme parallelism can be achieved on neuromorphic hardware. To demonstrate this, we implemented our recognition algorithm on the Loihi chip \cite{davies2018loihi,davies2021advancing}. The implementation consisted of three modules: template matching, binary decision, and recognition output [Fig.~\ref{Figure5}]. The first two modules were feed-forward spiking neural networks (SNNs) of leaky integrate-and-fire (LIF) neurons. These networks received onset and offset spikes from feature neurons in the CPU [Fig.~\ref{Figure1}]. The recognition module received input from the other two modules.

As in the CPU-based version, each template was embedded in the connections delays leaving feature neurons [Fig.~\ref{Figure5}(a), left]. These connections targeted a pair of LIF units representing template neurons. These neurons (here also referred to as ON and OFF neurons) integrate synchronized inputs and spike when at least 60\% of the template's spikes arrive aligned. For each syllable type, a two-compartment LIF neuron (plateau neuron) consisting of a dendrite and a soma was added to function as a conditional detector of OFF spikes given the prior arrival of ON spikes [Fig.~\ref{Figure5}(a), right]. The ON and OFF neurons of all templates of a type were connected to the dendritic and somatic compartments of a plateau neuron, respectively. When ON neurons spiked, the dendritic voltage rose above threshold and the neuron entered an UP state, which lasted while the dendritic voltage decayed back to threshold. If OFF neurons spiked during this interval, the soma spiked, signaling a detection. In its resting, or DOWN, state, the soma could not be driven above threshold by OFF inputs alone. A plateau neuron thus detected onset matches followed by offset matches, with the interval between them variable up to $\sim$217 ms when 60\% of the templates match \cite{jin2008decoding}, which exceeds the duration of Bengalese finch syllables. Each somatic spike inhibited the dendrite and returned the neuron to the DOWN state, so that a syllable elicited a single detection (Appendix~\ref{sec:methods:plateau_neurons}).
 
The binary decision module for a pair of syllables A and B consisted of a collision detection neuron (collision neuron for short) and two two-compartment neurons---one for each syllable [Fig.~\ref{Figure5}(b)]. The collision neuron received inputs from the ON units in the template modules of the two syllables. The input strengths were set such that the collision neuron spiked when both ON units spiked within $\sim$217 ms. A self-inhibitory mechanism prevented the collision neuron from spiking more than once. The collision neuron in turn was connected to the dendritic compartments of two output neurons [Fig.~\ref{Figure5}(b)]. When the collision neuron spiked, the output neurons transitioned to the UP state. The output neuron for syllable A received spikes from binary neurons that could distinguish A from B, while the output neuron for syllable B received spikes from neurons that could distinguish B from A. If these inputs arrived during the UP state, one of the output neurons fired, signaling the choice between the two types. We implemented binary decision modules for all pairs of syllable types for which binary neurons were trained. Thus, the responses of the output neurons were gated by the simultaneous detection of syllable pairs by template neurons. 

The recognition module consisted of LIF neurons, one per syllable type. These neurons had a long refractory period of $\sim$67 ms (20 timesteps at 300 Hz sampling rate), which prevented them from spiking more than once. Recognition neurons received inputs from both plateau and binary decision neurons [Fig.~\ref{Figure5}(c)]. Collision neurons provided feedback inhibition to the dendritic compartments of plateau neurons [Fig.~\ref{Figure6}], preventing them from spiking when ON spikes were triggered by more than one syllable type. In such cases, the spike of the recognition neuron was determined by the output of the binary module. Spikes generated by the recognition module represent the final predictions for syllable label and duration. Fig.~\ref{Figure5}(d) illustrates the typical input and output spike rasters during a continuous recognition task.

We compared the network’s performance to the CPU on the same recognition task of Fig.~\ref{Figure4}(a). WERs were comparable to the CPU results and decreased at a similar rate as the number of exemplars per syllable type increased [Fig.~\ref{Figure5}(e)]. Although WERs on Loihi were slightly higher, execution times were dramatically reduced [Fig.~\ref{Figure5}(f)]. As expected, parallel execution of all templates resulted in no change in execution times ($\sim$17 ms per annotated syllable, i.e., the total execution time on the test set divided by the number of annotated syllables) as the number of exemplars increased. For training sets containing 40 syllables per type, this corresponds to a speedup of two to three orders of magnitude. 

\subsection{Energy consumption}

We measured the energy consumption of the network with Loihi 1's energy probe \cite{davies2018loihi}. The network with 40 templates per syllable type processed the spikes computed on the CPU from 80 s of audio in 11.63 s, about 7 times faster than real time, and used 8.71 J. The dynamic power consumption was therefore 0.75 W. These figures cover the Loihi network only; they exclude the CPU computation of the feature-neuron and binary-neuron spikes.

\section{Discussion}
\label{sec:discussion}

Ultra-sparse spike representations of birdsong have been observed in higher auditory areas of the songbird brain \cite{schneider2013sparse}. Such representations are also robust to noise.
Feature-detecting neurons tuned to short snippets within auditory objects---such as birdsong syllables or words---transform a continuous auditory signal into a stream of ultra-sparse spikes \cite{schafer2014noise}.
We have shown that a template-matching method based on connection delays can detect syllable types from this spike stream.
Detection is further enhanced by binary feature-detecting neurons, whose spikes are highly discriminative when conditioned on conflicting simultaneous detections of multiple syllables by the template-matching process.
We implemented the algorithm on Intel’s neuromorphic chip, Loihi, achieving low latency that remained constant despite an increasing number of templates.
The success of our approach highlights the potential for low-latency and energy-efficient audio processing using ultra-sparse coding on neuromorphic devices.

A previous study applied the same coding scheme to the task of recognizing isolated spoken digits from the AURORA-2 database, and showed that performance was robust against challenging noise types, such as exhibition hall noise and babble noise, down to a 0 dB signal-to-noise ratio \cite{schafer2014noise}.
Isolated recognition involves only substitution errors (i.e., incorrect labels), whereas continuous recognition introduces additional challenges due to insertion and deletion errors.
We demonstrated continuous recognition using ultra-sparse spike coding for birdsong syllable recognition.
Compared to speech, birdsong is more manageable in some respects, as the syllables are stereotyped and the number of syllable types is limited.
However, the task remains non-trivial, as automatic birdsong recognition continues to pose significant challenges in birdsong research \cite{cohen2022automated,vengrovski2025tweetybert}.
Whether our approach can be successfully applied to continuous speech recognition remains an open question.
It is important to note that we have not incorporated syntactical information from birdsong syllables \cite{lu2025partially,vengrovski2025tweetybert}.
A complementary system that integrates syntactical structure could be valuable for extending this approach to speech recognition.

The binary feature-detecting neurons are an important component of our system.
These specialized neurons are sensitive to differences between specific syllable types.
As a result, they are highly informative for distinguishing those particular syllable pairs.
For other syllables, however, their activity may appear random or uninformative.
It would be interesting to investigate whether such neurons exist in the auditory system of the brain.

The spikes of the feature-detecting neurons and binary neurons were generated on CPUs.
To achieve a fully neuromorphic system, it is important to implement these neurons as spiking neural networks.
Spectral analysis on Loihi chips can already be performed using resonate-and-fire neurons \cite{shrestha2024efficient}.
It is plausible that the feature-detecting neurons could also be implemented using delay lines and synaptic weights driven by these spectral-analysis neurons.
Additionally, it would be interesting to explore whether such feature neurons could be tuned through unsupervised learning schemes, such as the Kohonen map \cite{kohonen2013essentials}.

Our template-matching scheme using axonal delays is closely related to a pioneering study that employed axonal delays to detect digits from continuous speech \cite{unnikrishnan1991connected}. In that earlier work, auditory signals were transformed into continuous neural activity representing high energy in specific frequency bands. These analog signals were directly summed after being delayed, and the weights were learned accordingly.
In contrast, our feature-detecting neurons utilize the full frequency information and emit spikes, which simplifies the template learning process. Additionally, transmitting spikes is inherently simpler than transmitting analog signals. Spike-based outputs are less susceptible to noisy fluctuations that can corrupt analog signals. Biologically, axons transmit spikes rather than analog signals, making our approach more plausible from a neurobiological standpoint.

Due to its simplicity, template matching was used in early automatic speech recognition (ASR) systems during the 1960s and 1970s \cite{juang2005automatic}. However, it soon became evident that this approach could easily exhaust computational resources as the number of templates increased, since the time required to search for matches grew accordingly.
Beginning in the 1980s, hidden Markov models (HMMs) took over and made ASR more practical \cite{juang2005automatic}. With advances in computing power, template matching has again become viable, although more sophisticated methods are now required to improve the speed of searching for matching templates \cite{de2007template}.
More recent approaches embed word templates into high-dimensional spaces using deep neural networks, enabling better discrimination through learning and more efficient search for matches \cite{kamper2016deep}.
We have demonstrated that neuromorphic devices enable the simultaneous evaluation of all templates, allowing fast responses even as the number of templates increases. This requires that the neuromorphic hardware can be scaled to accommodate the necessary number of templates. Given the rapid progress in neuromorphic technology \cite{kudithipudi2025neuromorphic}, it is conceivable that our approach could scale to support large-vocabulary speech recognition.

We have used onset and offset templates to detect the beginnings and ends of syllables. In the songs of some species, such as the canary, syllable durations can vary substantially due to features like harmonic stacks, which can exhibit large variability in duration \cite{cohen2022automated}.
The separation of onset and offset templates enables successful detection of such syllables without requiring explicit dynamic warping techniques \cite{tavares2022automatic}.
In the neuromorphic implementation, onsets and offsets are bridged by long-lasting plateau potentials in dendritic compartments. Indeed, neurons with dendritic compartments capable of generating plateau potentials can be wired to decode complex spike sequence patterns, even when the intervals between spikes vary over a wide range \cite{jin2008decoding}.
For speech recognition, it may be beneficial to use such neurons in combination with axonal delays to accommodate variability in speech tempo.

In conclusion, we have demonstrated that auditory objects, such as birdsong syllables, can be detected from ultra-sparse spike patterns generated by feature-detecting neurons driven by continuous auditory signals. We successfully implemented this process on Loihi chips, highlighting the potential of neuromorphic devices for low-latency and low-energy processing of complex signals such as speech.

\begin{acknowledgments}
We thank Marta Kanne for testing the software and helping annotate songs.
Research supported by NSF award EF-1822476 and a gift fund from Intel Corporation for neuromorphic research. The funders had no role in
study design, data collection and analysis, decision to publish, or preparation of the
manuscript.
\end{acknowledgments}

\appendix

\section{Methods}
\label{sec:methods}

\subsection{Datasets}
\label{sec:datasets}

Undirected songs of 8 adult Bengalese finches ($>$120 days post hatch) were recorded at McGill University. Birds were housed individually in sound-attenuating chambers (“soundbox”; TRA Acoustics, Ontario, Canada) and recorded using an omnidirectional microphone (Countryman Associates, CA, USA). Sound Analysis Pro 2011 (SAP; \url{http://soundanalysispro.com}) was used to detect and digitize songs at 44.1 kHz. All procedures were approved by the McGill University Animal Care and Use Committee (\#7149) in accordance with the guidelines of the Canadian Council on Animal Care. Songs from an additional group of 11 adult Bengalese finches were selected from a publicly available dataset and included as an additional benchmark \cite{koumura2016automatic}.

\subsection{Song annotation}
\label{sec:song_annotation}

We designed a semi-automated pipeline to annotate the recordings of 8 Bengalese finches [see Table~\ref{tab:datasets}]. We initially manually annotated a subset of the recordings from each bird, ensuring that all syllable types were identified and labeled. We then trained our recognition system and used it to automatically annotate approximately 10 additional recordings, each roughly 30 seconds long. We manually curated the results by correcting mislabeled syllables and adjusting syllable boundaries. We repeated this process iteratively, each time using a larger training set, until all recordings were annotated.

\begin{table}[t]
  \caption{Dataset IDs, number of syllable types, test set sizes, and number of syllables available for training. Recording files were divided into a test set (60\%) and a training pool (40\%) from which training sets were drawn. All test syllables were used in the benchmarks. The first eight birds are from this study, while the remaining eleven are from Ref.~\cite{koumura2016automatic}.}
  
  \label{tab:datasets}
  \centering
  \begin{tabular}{|c|cccc|}
  \hline
  ID         & \# types & \# test  & \# train & \# total \\
  \hline
  s933\_WOtutor & 19    & 10221 & 6478  & 16699 \\
  w27o185\_15mo & 13    & 16214 & 10273 & 26487 \\
  w31o179\_15mo & 9     & 16736 & 10721 & 27457 \\  
  w34o164\_4mo  & 10    & 21762 & 14380 & 36142 \\
  w72o54\_06mo  & 12    & 10824 & 7446  & 18270 \\
  w74o56\_06mo  & 9     & 14553 & 9103  & 23656 \\
  w96o142\_06mo & 13    & 14404 & 9982  & 24386 \\
  w100o80\_06mo & 18    & 9659  & 5820  & 15479 \\
  Bird0         & 9     & 4602  & 3116  & 7718  \\
  Bird1         & 12    & 21231 & 14499 & 35730 \\
  Bird2         & 17    & 15803 & 10324 & 26127 \\
  Bird3         & 9     & 17631 & 11839 & 29470 \\
  Bird4         & 5     & 16152 & 10739 & 26891 \\
  Bird5         & 7     & 12165 & 8360  & 20525 \\
  Bird6         & 5     & 10493 & 7160  & 17653 \\
  Bird7         & 7     & 12287 & 8435  & 20722 \\
  Bird8         & 4     & 3070  & 1915  & 4985  \\
  Bird9         & 6     & 12035 & 7506  & 19541 \\
  Bird10        & 12    & 3725  & 2018  & 5743  \\
  \hline
  \end{tabular}
\end{table}

\subsection{Feature processing}
\label{sec:methods:feature_processing}

We devised a preprocessing pipeline inspired by models of the auditory periphery. Recordings were first band-pass filtered between 1 and 10 kHz with a fifth-order Butterworth filter, applied forward and backward to avoid phase distortion. They were then passed through a bank of 32 fourth-order gammatone filters, which simulate the spectral analysis performed by the cochlea \cite{patterson1987efficient}. The impulse response of a fourth-order gammatone filter with center frequency $f$ is
\begin{equation}
	g(f, t) = g_0 t^3 \exp{\{-2\pi b(f)t\}} \cos{\left(2\pi f t\right)},
\end{equation}
where $g_0$ sets the gain and $b(f)$ the bandwidth. Our implementation followed Ref.~\cite{slaney1993efficient}, but used an exact calculation of the filter coefficients, with the gain normalized to unity at the center frequency.

Bandwidths were set by an equivalent rectangular bandwidth (ERB) function \cite{glasberg1990derivation}. Although songbirds can hear frequencies up to 10 kHz, their hearing is most sensitive between 2 and 5 kHz \cite{woolley2013songbird}. To emphasize this range, we replaced the human ERB function with a linear function fitted by least squares to the starling data of Ref.~\cite{marean1998auditory}:
\begin{equation}
	\mathrm{ERB}(f) = f/Q+f_{\mathrm{mBW}},
\end{equation}
where $Q=10.49$ is the (dimensionless) quality factor and $f_{\mathrm{mBW}}=88.29$~Hz is the minimum bandwidth. These values are similar to those suggested in Ref.~\cite{moore1983suggested} for human hearing at low to moderate frequencies and levels. Following Ref.~\cite{holdsworth1988implementing}, we set $b(f)=1.019 \times \mathrm{ERB}(f)$. The 32 center frequencies were equally spaced between 1 and 10 kHz on the ERB-number scale, making filters denser at lower frequencies where they are also narrower.

The filter outputs were converted into spectrograms in five steps. (i) Each channel was advanced in time by $3/(2\pi b(f)$), rounded down to a whole number of samples, to compensate for the longer delays of the low-frequency filters. (ii) The outputs were squared to obtain the instantaneous power in each band, half-wave rectified, and compressed with a power of 0.01 to reduce their dynamic range. (iii) The envelope of each channel was extracted by locating the maxima of the compressed power within consecutive 5 ms windows and interpolating linearly between them. (iv) The envelopes were resampled to 300 Hz. (v) The envelope of each channel was standardized by $(e_i-\mu_i)/\sigma_i$, where $i$ indexes the channel, $e_i$ is the envelope and $\mu_i$ and $\sigma_i$ are the mean and standard deviation of channel $i$ computed using all the recordings. 

Feature vectors were formed by stacking 11 consecutive spectrogram frames, giving $\sim$37 ms segments represented as 352-dimensional vectors (11 frames $\times$ 32 channels). Each vector was divided by its Euclidean norm, with a small constant ($10^{-8}$) added to the norm to avoid division by zero. Feature vectors are therefore dimensionless and insensitive to the overall sound level. Sliding the window forward one frame at a time produced the trajectories in the feature space, as shown in Fig.~\ref{Figure1}(a).

To speed up decoding, we discarded low-power audio segments before computing the responses of feature neurons. The instantaneous power $P(t)$ was computed directly from the waveform with a 20 ms sliding Kaiser window ($\pi \alpha = 14$) advanced by one spectrogram frame (3.3 ms). In each window, the samples were squared and the maximum was taken. The result was smoothed with a Gaussian filter with a standard deviation of one frame, half-wave rectified, and compressed with a power of 0.01. Segments were discarded where $P(t)$ fell below the threshold
\begin{equation}
  P_{\mathrm{min}} = \text{MED}(P)-\text{MAD}(P),
\end{equation}
where $\text{MED}$ and $\text{MAD}$ are the median and the median absolute deviation of $P$, computed separately for each recording. To avoid cutting syllables apart, we merged adjacent segments separated by less than 100 ms. We also discarded isolated segments shorter than 20 ms, which is shorter than any syllable we analyzed.

\subsection{Feature neurons}
\label{section:methods:feature_neurons}

Feature neurons were trained by nonlinear (Gaussian kernel) support vector machines (SVMs) \cite{cortes1995support} using the Python interface to LIBSVM in Scikit-Learn \cite{chang2011libsvm,scikitlearn2011}.  We effectively used hard margins by setting the soft-margin penalty to $C=10^{20}$. The response of a neuron to the time-dependent input ${\bf x}(t)$ is the SVM decision function evaluated at time $t$ [Eq.~\eqref{equation:feature_detector_response}]. Because feature vectors were normalized (Appendix~\ref{sec:methods:feature_processing}), $V(t)$ in Eq.~\eqref{equation:feature_detector_response} is dimensionless. 

A feature neuron detected a feature whenever $V(t)$ rose above the decision boundary ($V(t)=0$). Spike times were defined as the time when $V(t)=-1+\delta$, with $\delta=0.2$, just above the SVM lower margin $V=-1$. The onset spike was assigned to the nearest preceding time at which $V(t)$ crossed this level from below, and the offset spike to the nearest following time at which it crossed from above. Onset and offset spikes therefore mark the times when the input trajectory enters and leaves the neighborhood of the neuron's decision region.

Feature neurons were trained one time bin at a time from reference syllables, with a $K$-fold scheme. We partitioned the training syllables at random into $K$ groups of equal size, each containing the same number of syllables per type. Each group in turn served as the held-out set, and the remaining $K-1$ groups as the training set. We used $K=N/5$, i.e., $K=2,4,8$ for $N=10,20,40$ training syllables per type; for $N=5$, all syllables were used for training. In each fold, we randomly selected one reference syllable $S_r$ per type and, for each time bin $\tau$ of $S_r$, trained one feature neuron as follows.
\begin{enumerate}
\item The positive training set (target $y=1$) consisted of the reference feature vector $S_r(\tau)$ and, from each of the other training syllables of the same type, the feature vector with the highest cross-correlation with $S_r(\tau)$;
\item The negative training set (target $y=-1$) consisted of all the feature vectors of 5 syllables sampled at random from each of the other syllable types. To make neurons robust to background events and reduce false alarm rates, we added feature vectors from audio segments that did not contain annotated syllables (silence, calls, and background noise);
\item Because hard-margin SVMs can overfit when forced to separate similar vectors, we removed from the negative set any vector that cross-correlated by more than 0.97 with any positive vector. If SVM optimization failed to converge, this threshold was lowered in steps of 0.05, up to 10 times. Lowering the threshold smoothed $V(t)$ and reduced the number of false positive spikes;
\item The kernel parameter $\gamma$ in Eq.~\eqref{equation:feature_detector_response}, which sets the width of the decision region around each support vector, was set to $\gamma=2/d_{\min}^2$, where $d_{\min}$ is the smallest Euclidean distance between $S_r(\tau)$ and the other positive vectors.
\end{enumerate}
Because neighboring time bins can have similar acoustics, this procedure can produce several redundant neurons. To prevent the system from scaling unnecessarily, we removed poorly performing or highly correlated neurons. Each neuron was scored on its held-out set as an isolated-syllable detector: its error rate was the fraction of held-out syllables of its own type to which it did not respond, plus those of other types to which it did. For $N=20$ and $N=40$, we then clustered the neurons of each syllable type. For $M$ neurons of a given type, we computed the correlation matrix of their responses $V(t)$ to the training syllables, using $M/2$ clusters for $N=20$ and $M/4$ for $N=40$, and performed spectral clustering on this matrix. In each cluster we kept the 4 neurons with the lowest error rates and discarded the rest.

\subsection{Template training and matching on the CPU}
\label{sec:methods:template_training}

A template consisted of the onset and offset spike times produced by the feature neurons in response to a syllable. A feature neuron was included in a template only if it spiked during the template's syllable, i.e. if $V(t)$ reached the decision boundary ($V(t)=0$) at least once. Neurons included in this way were kept in fixed order, guaranteeing that templates stored the identity of each neuron as well as its spike times. Spike times were stored as delays relative to a reference neuron: onsets relative to the onset of the first feature neuron, and offsets relative to the offset of the last. ON template neurons were therefore anchored to the beginning of the syllable and OFF template neurons to the end.

Templates were then used to train the ON and OFF neurons for onsets and offsets, respectively. For each template $T_1$ of a syllable type, we shifted the spikes of every template $T_2$ of the same type, including $T_1$ itself, by the delays in $T_1$. When $T_2$ resembles $T_1$, all spikes become temporally aligned. The aligned spikes were converted into binary matrices (rows representing feature neurons, columns representing time, and ones representing spikes) and convolved in time with a Gaussian kernel of standard deviation 5 ms to tolerate jitter in spike timing. From each aligned template we extracted one temporal window ${\bf s}(t)$, centered on its mean aligned spike time, and these windows formed the positive training set. The negative set consisted of three parts:
\begin{enumerate}
\item The spike responses to training syllables of other types that made at least one of $T_1$'s neurons spike, aligned with the delays of $T_1$, from which all windows were taken;
\item The all-zero pattern, representing the absence of input;
\item For templates with $M>2$ neurons, $10\times M$ random patterns in which each neuron spiked independently with probability $p=0.01/\{1+\exp[-0.1(M-5)]\}$.
\end{enumerate}
The negative patterns were smoothed with the same Gaussian kernel. The window duration was set for each syllable type to three times the standard deviation of the aligned spike times across all pairs of its templates, bounded between 17 ms and 100 ms (5 and 30 frames). The windows were vectorized as in Appendix~\ref{sec:methods:feature_processing} and used to train linear SVMs. The linear SVM decision function can be recast from Eq.~\eqref{equation:feature_detector_response} as a sum of $M$ linear filtering operations, one for each feature neuron:
\begin{equation}
  V^{\mathrm{temp}}(t) = \sum_{i=1}^M \left( w_i * s_i \right)(t)+V_0^{\mathrm{temp}}.
\label{equation:template_response}
\end{equation}
Here, $s_i$ is the smoothed spike train of the $i$-th feature neuron, $w_i$ is the $i$-th row of the SVM weights arranged as a matrix (neurons $\times$ time), and $V_0^{\mathrm{temp}}$ is the offset [see Figs.~\ref{Figure2}(c)--\ref{Figure2}(d)]. Feature neurons that fired reliably across the training syllables received large weights. ON and OFF spike times were computed as the peak of $V^{\mathrm{temp}}(t)$ once it crossed the SVM decision threshold, similar to the implementation of feature neurons (Appendix~\ref{section:methods:feature_neurons}).

During recognition, onset (or offset) spikes of feature neurons in each template were shifted by the template's delays, smoothed, and passed through Eq.~\eqref{equation:template_response} to obtain ON (or OFF) spikes.

\subsection{Binary neurons}
\label{sec:methods:binary_neurons}

For an ordered pair of syllable types (B versus A), we trained an ensemble of binary neurons with AdaBoost \cite{freund1997decision}, following the face detection algorithm of Ref.~\cite{viola2004robust}. AdaBoost combines weak classifiers into a strong one. Here, each weak classifier is a binary neuron that examines a small rectangular spectrotemporal patch of the feature vector ${\bf x}(t)$ and discriminates between the two syllable types only. Its response is linear, $V^{\mathrm{bin}}_i({\bf x}(t)) = {\bf w}_i \cdot {\bf x}(t) + V_{0,i}^{\mathrm{bin}}$, where ${\bf w}_i$ is nonzero only within the neuron's patch, which keeps it weak enough for boosting. When applied to a syllable or segment, neuron $i$ returns $h_i=+1$ (type B) if $V^{\mathrm{bin}}_i({\bf x}(t))>0$ at any time $t$ within it, and $h_i=-1$ (type A) otherwise. The ensemble response is the normalized weighted vote
\begin{equation}
  V^{\mathrm{bin}} = \frac{\sum_{i=1}^T \omega_i h_i}{\sum_{i=1}^T \omega_i},
  \label{eq:binary_response}
\end{equation}
where $\omega_i$ are the weights learned by AdaBoost and $T=10$ is the number of binary neurons per ensemble.

Each binary neuron was trained with stochastic gradient descent to minimize the hinge loss with an L2 regularization term of strength $0.001$ (\texttt{SGDClassifier} in Scikit-Learn \cite{scikitlearn2011}). Training stopped when the loss failed to decrease by at least $10^{-8}$ for 5 consecutive epochs.

To train an ensemble, we labeled the training syllables of the two types $y_k=+1$ (B) or $y_k=-1$ (A) and assigned them initial weights $p_k$, so that each type carried half of the total weight: $p_k=1/(2N_{\mathrm{B}})$ or $1/(2N_{\mathrm{A}})$, where $N_{\mathrm{B}}$ and $N_{\mathrm{A}}$ are the numbers of training syllables of each type. We then iterated the following steps for $t=1 \ldots T$ \cite{viola2004robust}:
\begin{enumerate}
  \item Train 20 candidate binary neurons. For each candidate:
    \begin{enumerate}
      \item Sample 5 syllables of type B and 5 of type A at random, with probabilities proportional to $p_k$;      
      \item Choose a random time in a random type B syllable, and a random rectangular patch within the $\sim$35 ms window ending there. The patch spans between 13 and 33 ms and at least 8 of the 32 frequency bands;
      \item From each of the sampled type B syllables, take the window with the highest cross-correlation with the chosen window, keeping it if the correlation exceeds 0.5. The patches of these windows, together with the chosen one, form the positive set;
      \item From the sampled type A syllables, take the same patch from all windows. These form the negative set;
      \item Train the binary neuron to separate the positive set from the negative set;
    \end{enumerate}
  \item Evaluate every candidate on all training syllables of the two types and keep the one with the smallest weighted error $\epsilon_t = \sum_{k:\,h_t(k) \neq y_k} p_k$. Add it to the ensemble with weight $\omega_t=\frac{1}{2}\ln[(1-\epsilon_t)/\epsilon_t]$. In practice, a constant of 0.01 was added to the numerator and to the denominator to avoid division by zero;
  \item Update the syllable weights, $p_k \leftarrow p_k \exp\{-\omega_t y_k h_t(k)\}$, which increases the weights of misclassified syllables, and normalize, $p_k \leftarrow p_k/\sum_j p_j$.
\end{enumerate}

\subsection{Template recognition on the CPU}
\label{sec:methods:template_matching}

Recognition proceeded in three stages. First, we computed the onset and offset spikes of feature neurons for the pre-segmented audio (Appendix~\ref{sec:methods:feature_processing}) and the ON and OFF spikes of template neurons (Appendix~\ref{sec:methods:template_training}). To save computation, a syllable type was considered in the recording only if any of its feature neurons were active there. 

Second, the ON and OFF spikes of each template were paired into candidate segments, each consisting of an ON time, an OFF time, and the template's syllable type. From each ON spike, we scanned forward in time and took the first OFF spike of the same template within $\langle T_i\rangle+3\sigma_{T_i}+40$~ms, where $\langle T_i\rangle$ and $\sigma_{T_i}$ are the mean and standard deviation of the durations of syllable type $i$ in the training set. ON spikes without a matching OFF spike were discarded. Each segment was assigned a score $e_i=V^{\mathrm{temp}}_i(\text{ON}_i)+V^{\mathrm{temp}}_i(\text{OFF}_i)$, where $V^{\mathrm{temp}}_i$ is the response of the ON or OFF template neuron [Eq.~\eqref{equation:template_response}] at the segment boundary.

Third, because each syllable type has many templates, a single syllable typically produces several overlapping segments, sometimes with different labels. Two segments $i$ and $j$ were considered to collide if one contained the other, or if their overlap exceeded a fraction $f$ of the shorter segment:
\begin{equation}
\begin{split}
  &|[\text{ON}_i, \text{OFF}_i] \cap [\text{ON}_j, \text{OFF}_j]| \\
  &\quad \geq f \min\{ \text{OFF}_i-\text{ON}_i, \text{OFF}_j-\text{ON}_j \}.
\end{split}
\end{equation}
Collisions were resolved in three passes, using a different overlap fraction for each:
\begin{enumerate}
\item Segments of the same syllable type ($f=0.75$): the segment with the higher score $e_i$ was kept;
\item Segments of different types ($f=0.25$): if the scores differed by more than 0.25, the segment with the lower score was removed;
\item Segments of different types still colliding ($f=0.4$): the binary neuron ensembles for the two types (Appendix~\ref{sec:methods:binary_neurons}) were applied to the feature vectors from the earlier ON time to the later OFF time. When ensembles for both orderings of the pair (B versus A and A versus B) existed, the segment whose type received the higher ensemble response $V^{\mathrm{bin}}$ [Eq.~\eqref{eq:binary_response}] was kept. When only one existed, the sign of its response decided.
\end{enumerate}
Within each pass, colliding pairs were processed in order of decreasing score difference, and pairs involving an already removed segment were skipped. Binary neuron ensembles were trained on demand: when two syllable types collided in the third pass and no ensemble existed for them, ensembles for both orderings of the pair were trained from the training syllables of the two types, and the pass was repeated.

\subsection{Plateau neurons}
\label{sec:methods:plateau_neurons}

We used a model inspired by dendritic plateau neurons \cite{jin2008decoding} to detect a template onset match followed by an offset match. Each plateau neuron was a Loihi 1 multi-compartment neuron with two compartments, a dendrite and a soma \cite{davies2018loihi}. Each compartment followed Loihi's discrete-time leaky integrate-and-fire dynamics,
\begin{eqnarray}
  u[t] &=& (1-\lambda_u)\,u[t-1] + \sigma[t],  \\
  v[t] &=& (1-\lambda_v)\,v[t-1] + u[t],
\end{eqnarray}
where $u$ is the synaptic current, $v$ the voltage, $\sigma[t]$ the weighted sum of input spikes at timestep $t$, and $\lambda_u, \lambda_v \in [0,1]$ the current and voltage decay rates. Loihi specifies these rates as integers $d \in [0, 4096]$ with $\lambda = d/4096$, giving a time constant $\tau \approx 4096/d$ timesteps. One timestep corresponded to one spectrogram frame (3.33 ms). Throughout the network we set the current decay to 4096 ($\lambda_u = 1$), so the current carries no memory and $v[t] = (1-\lambda_v)\,v[t-1] + \sigma[t]$.

The two compartments were joined so that the dendrite never spiked and the soma could spike only when both compartments were above their thresholds, $\theta_\mathrm{d}$ and $\theta_\mathrm{s}$. The neuron was in the UP state while $v_\mathrm{d} > \theta_\mathrm{d}$ and in the DOWN state otherwise. In the DOWN state the soma integrated its inputs but could not spike. There was one plateau neuron per syllable type. Its dendrite received the ON spikes of all templates of that type, and its soma received their OFF spikes.

The order of the two events was enforced by the somatic decay. The soma was given a strong voltage decay, much stronger than that of the dendrite, so that its voltage remained above threshold only briefly after a burst of OFF spikes. A somatic spike was therefore possible only if the dendrite was already in the UP state or crossed its threshold at nearly the same time, that is, if the ON matches preceded the OFF matches. When the soma fired, its voltage reset but the dendritic voltage did not, so later OFF spikes from the same syllable could have triggered a second detection. We therefore added an inhibitory connection from the soma to the dendrite, which returned the neuron to the DOWN state after each detection.

The UP state lasted until the dendritic voltage decayed back to threshold. If $n$ ON template neurons spike simultaneously, each with weight $w_\mathrm{d}$, the UP state lasts
\begin{equation}
  \Delta t = \tau_\mathrm{d} \ln\!\left( \frac{n w_\mathrm{d}}{\theta_\mathrm{d}} \right)
  \label{eq:up_duration}
\end{equation}
timesteps. $\Delta t$ must exceed the duration of the longest syllable type. However, it also grows with the number of matching templates $n$, which is unrelated to syllable duration. Since $d\Delta t/dn = \tau_\mathrm{d}/n$, we reduced this dependence by using the shortest dendritic time constant compatible with Loihi's weight range ($w_\mathrm{d} \in [1, 255]$): a voltage decay of 272 ($\tau_\mathrm{d} \approx 15$ timesteps, 50 ms) and $\theta_\mathrm{d} = 10$. For a system with $N$ templates per syllable type, the dendritic weight was set by inverting Eq.~\eqref{eq:up_duration}, so that the UP state lasts $\Delta t = 65$ timesteps (217 ms) when 60\% of the templates ($n=\lfloor 0.6N \rfloor$) match. Because Loihi stores voltages as integers, the actual UP durations deviate slightly from Eq.~\eqref{eq:up_duration}.

\subsection{Template recognition on a neuromorphic device}
\label{sec:methods:tmn_template_layer}

\begin{figure*}[!tp]
\includegraphics[width=\textwidth]{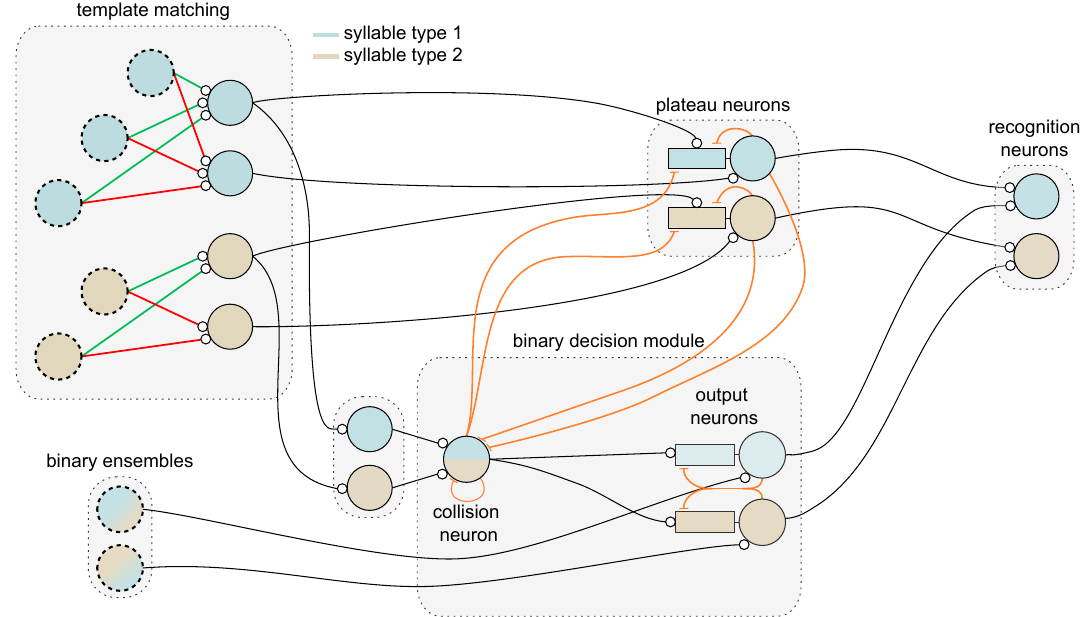}
\caption{
{\bf Full diagram of the spiking neural network implemented on the Loihi chip}. Inhibitory connections are shown in orange. Excitatory connections are shown in black. Template onset and offset delay lines are shown in green and red, respectively.
}
\label{Figure6}
\end{figure*}

The complete network is shown in Fig.~\ref{Figure6}. It received three spike streams computed on the CPU: the onset and offset spikes of the feature neurons and the spikes of the binary neurons. One network timestep corresponded to one spectrogram frame (3.3 ms). Unless stated otherwise, all neurons were single-compartment LIF neurons with instant current decay (Appendix~\ref{sec:methods:plateau_neurons}).

{\it Template matching.} Each template was encoded by one ON and one OFF template neuron. Onset spikes of the feature neurons in the template were sent to the ON neuron, and offset spikes to the OFF neuron, through connections whose delays aligned the spikes as in the CPU implementation [Fig.~\ref{Figure5}(a)]. Loihi 1 stores synaptic delays in 6 bits, limiting them to 62 timesteps. For systems with longer template delays, we compressed time by multiplying all delays and input spike times by $62/D_{\max}$, where $D_{\max}$ is the largest onset template delay, and rescaled the output spike times accordingly. Because templates differ widely in the number of spikes they contain, the threshold of each template neuron was normalized to its template size: $\theta = \max[1, \lfloor f_{\mathrm{th}} n_F w \rfloor]$, where $n_F$ is the number of feature neurons in the template, $w$ the synaptic weight, and $f_{\mathrm{th}} = 0.6$ the fraction of spikes that must arrive aligned. The voltage decay of the template neurons, which sets how much timing jitter is tolerated, was tuned by trial and error. A refractory period ensured that each template neuron spiked at most once per syllable.

{\it Syllable detection.} The ON and OFF neurons of all templates of a syllable type were connected to the dendrite and soma, respectively, of a single plateau neuron for that type (Appendix~\ref{sec:methods:plateau_neurons}). A somatic spike signaled detection of the syllable. Since the delays align the onset spikes to the end of the template, offset spikes were given an additional delay equal to the largest onset template delay, which kept the ON and OFF streams synchronized.

{\it Collision detection.} Because the dendrite of a Loihi plateau neuron cannot spike, the onset of a detection is not directly available. We therefore added one coincidence-detection neuron per syllable type. It received the same ON spikes as the dendrite of the plateau neuron and had the same parameters, except that it spiked and reset when its threshold was crossed. Its refractory period ensured a single spike per syllable onset. For each pair of syllable types with binary neurons, a collision neuron received input from the two coincidence-detection neurons, through a small uniform delay. Its weights, threshold, and decay were set so that it spiked when both inputs arrived within a short window, but not when two spikes of the same type arrived one refractory period apart. An inhibitory self-connection prevented each collision neuron from spiking more than once per collision. Two further inhibitory pathways kept collision neurons from firing falsely. Each plateau neuron, on detecting its syllable, inhibited the collision neurons of that type, so that adjacent or repeated syllables did not trigger a collision; the small delay ensured this inhibition arrived in time. Collision neurons that share a syllable type inhibited each other, so that resolving one collision cleared the others primed by the same onset.

{\it Binary decision.} When a collision neuron spiked, it inhibited the dendrites of the plateau neurons of both syllable types, returning them to the DOWN state so that neither could report a detection. At the same time, it excited the dendrites of two output neurons, one per syllable type, putting them in the UP state [Fig.~\ref{Figure5}(b)]. Output neurons were two-compartment neurons like the plateau neurons. The soma of the output neuron for type A received the spikes of the binary neurons discriminating A from B, and vice versa. The binary neuron spikes were delayed by the largest onset template delay to stay synchronized with the template stream. Whichever output neuron fired first won: its spike inhibited the dendrites of both output neurons, returning the pair to the DOWN state. This first-spike competition requires ensembles for both orderings of each pair (A versus B and B versus A).

{\it Recognition.} The recognition module contained one LIF neuron per syllable type, which spiked on any input from the plateau neuron or output neurons of its type [Fig.~\ref{Figure5}(c)]. A refractory period of 20 timesteps ($\sim$67 ms) prevented double reports when both pathways responded to the same syllable. Each recognition spike marks the end of a syllable. The total delay of the network is the largest onset template delay plus two timesteps, independent of the number of templates; output spike times were shifted back by this amount. Because the network reports only syllable ends, syllable onsets were estimated by subtracting the mean duration of the recognized type, computed from the training set.

\subsection{Benchmarks}
\label{sec:methods:benchmark}

For each bird, the recordings were split at the file level: 60\% of the files, selected at random with a fixed seed, formed the test set, and the remaining 40\% formed the training pool. The same split was used for every condition and for both the CPU and neuromorphic systems. For each condition, $N = 5, 10, 20$, or 40 syllables per type, we drew 10 training sets independently from the training pool by sampling $N$ syllables of each type at random without replacement. Syllables of the training pool that were not sampled were discarded. Each training set also included approximately 60~s of unannotated audio from the training recordings, representing silence and background noise, which was used to build the negative sets of the feature neurons (Appendix~\ref{section:methods:feature_neurons}). One system was trained on each training set, and all systems were tested on the full test set. CPU and neuromorphic systems were trained and tested on the same sets (data from this paper). Table~\ref{tab:datasets} summarizes the sizes of the benchmarks.

Predictions were scored against the manual annotations with the word error rate (WER) [Eq.~\eqref{equation:wer}]. For each test recording, the predicted and annotated syllables were ordered by time into label sequences, which were aligned by minimum edit distance; each alignment step counts as a correct label, a substitution, a deletion, or an insertion. The counts were summed over all test recordings, so that $K$ in Eq.~\eqref{equation:wer} is the total number of annotated test syllables. Annotations of calls and noise, which the systems were not trained to recognize, were excluded from scoring, together with any prediction overlapping such an annotation by at least 10\% of the shorter of the two. Because the alignment uses only the order of the labels, the WER does not penalize errors in the predicted syllable boundaries.

%\nolinenumbers

%\bibliographystyle{jneurosci}
% The bibtex filename

\bibliography{Auditory.bib}

\end{document}